\documentclass[aps,prb,superscriptaddress,twocolumn,floatfix]{revtex4}

\usepackage{graphicx,color}
\usepackage{dcolumn}
\usepackage{bm}
\usepackage{stmaryrd}
\usepackage{latexsym}
\usepackage{amssymb}
\usepackage{amsfonts}
\usepackage{amsmath}

\begin{document}

\title{First-principles study of an organometallic S=1/2 kagom\'{e} compound}

\author{Zheng Liu}
\affiliation{Institute for Advanced Study, Tsinghua University, Beijing 100084, China}
\affiliation{Department of Materials Science and Engineering, University of Utah, Salt Lake City, UT 84112, USA}

\author{Jia-Wei Mei}
\affiliation{Perimeter Institute for Theoretical Physics, Waterloo, Ontario, N2L 2Y5 Canada}

\author{Feng Liu}
\affiliation{Department of Materials Science and Engineering, University of Utah, Salt Lake City, UT 84112, USA}
\affiliation{Collaborative Innovation Center of Quantum Matter, Beijing 100084, China}

\date{\today}

\pacs{}

\begin{abstract}
Cu(1,3-benzenedicarboxylate) [Cu(1,3-bdc)] contains structurally perfect kagom\'{e} planes formed by Cu$^{2+}$ ions without the presence of diamagnetic defects. This organometallic compound should have served as a precious platform to explore quantum frustrated magnetism, yet the experimental results so far are mysterious, leading to questions such as ``Is Cu(1,3-bdc) just a trivial weak ferromagnet?''. Using the the density functional theory, we have systematically studied the electronic and magnetic properties of Cu(1,3-bdc), putting forth a theoretical basis to clarify this novel material. We present numerical evidence of a dominating antiferromagnetic (AFM) exchange between nearest-neighbor (NN) Cu$^{2+}$ as experimentally extracted from the high-temperature susceptibility data. We further show that beyond the NN AFM exchange, the additional interactions in Cu(1,3-bdc) have similar strength as those in the well-studied kagom\'{e} antiferromagnet, Herbertsmithite, by designing a comparative study. In the end, we discuss our understanding on the phase transition and FM signals observed under low temperature.
\end{abstract}

\maketitle

\section{Introduction}

Since Anderson's proposal in the 70s \cite{Anderson73RVB}, the concept of quantum spin liquids (QSLs) has now become an indispensable brick laid upon the two milestones of modern condensed matter physics, namely the high-temperature cuprate superconductivity and the fractional quantum Hall effect \cite{XiaogangWen}. Experimental discovery of QSLs in the so-called quantum frustrated materials is a long-sought goal to bring out exotic new quasi-particles and gauge fields never encountered before \cite{Book11FrustMag}.

After decades of searching, several promising examples have now emerged \cite{Science08PALeeQSL,Nature10BalentsQSL}. The hottest candidate at present is perhaps Herbertsmithite ZnCu$_3$(OH)$_6$Cl$_2$, which realizes the $S=\frac{1}{2}$ AFM Heisenberg model on the 2D kagom\'{e} lattice \cite{Nocera05Herbert}. Extensive theoretical studies have suggested that this model is likely to achieve a QSL ground-state, despite close in energy with other competing phases \cite{Sachdev92kagomedisorder,PRB97disorderkagome,Matthew07kagomeliquid,RanWen07Kagome,Yan11DMRGkagome,PRB13Yasir1,PRB13Yasir2}. Experiments on Herbertsmithite have also shown QSL-like features, such as the absence of any observed magnetic order down to 50 mK \cite{YLee07Neutron,Mendel07HerbertUSR} and an unusual continuum of spin excitations \cite{YLee12Neutron}. However, the inevitable Cu/Zn substitutional defects make the interpretation of experimental data difficult \cite{Harrison08NMRHerbert}. It remains an open debate whether these defects obscure the intrinsic signals under low temperature \cite{Singh07suscept}.

Cu(1,3-bdc), synthesized in the same group three years after Herbertsmithite, also features with structurally perfect Cu$^{2+}$ kagom\'{e} planes \cite{JACS08CuBDC}. A great advantage of Cu(1,3-bdc) is that the substitutional defects are automatically avoided. Unfortunately, Cu(1,3-bdc) has been found to undergo a phase transition at $T_c\sim2K$ \cite{JACS08CuBDC,Keren09usR}, which appears to exclude the possibility of a QSL ground state. This material has thus been largely overlooked. However, given the structural similarity between Cu(1,3-bdc) and Herbertsmithite, a natural question is why the spins behave so differently in these two materials. It is desirable to better understand the electronic properties of Cu(1,3-bdc), as it would in turn help to understand the QSL-like behaviors of Herbertsmithite, and  further reveal key factors to achieve QSLs.

An overview on the experimental data of Cu(1,3-bdc) shows puzzling ambiguities. Fitting the high-temperature susceptibility data to the Curie-Weiss law yields a Weiss constant $\theta=-33K$, suggesting a mean nearest-neighbor (NN) AFM exchange $J_1\sim30K$ \cite{JACS08CuBDC}. The ratio $|J_1|/T_c>10$ indicates a strong frustration effect. Later, muon spin relaxation ($\mu$SR) observes persistent spin fluctuation below $T_c$, which further supports the frustration scenario \cite{Keren09usR}. On the other hand, the magnetization data around $T_c$ displays a ferromagnetic (FM)-like curve \cite{Keren14PRBCuBDC} and a small hysteresis loop with the coercive field of 10.5 Oe \cite{JACS08CuBDC}. A recent work starting from the FM hypothesis extracted a mean NN FM exchange $J_1\sim-2K$ from the electron spin resonance lineshape, proposing Cu(1,3-bdc) be rather a weak ferromagnet without frustration ($|J_1|/T_c\sim1$) \cite{Keren14PRBCuBDC}. In addition, there is an unpublished neutron scattering work, which employs this FM scenario to interpret the dynamic structure factor \cite{Robin14APS}.

This Article aims to provide a first-principles description of Cu(1,3-bdc) based on density functional theory (DFT) \cite{RMP89DFT} and possibly resolve some lasting controversies. Our primary goal is to determine the type of NN spin exchange in order to rationalize Cu(1,3-bdc) as a kagom\'{e} antiferromagnet. The second goal is to characterize additional interactions in this material, such as the longer-range spin exchange and Dzyaloshinskii-Moriya (DM) interaction, in order to explain the subtleties in the experimental data. In Sec. \ref{sec:method}, we describe the general formalism of our calculation. In Sec. \ref{sec:band}, we show the structural and single-electron properties of Cu(1,3-bdc). Sections \ref{sec:wannier} and \ref{sec:+U} present the results from Wannier function analysis and DFT+U total energy calculation, respectively. Section \ref{sec:soc} incorporates spin-orbit coupling (SOC) into the calculation and estimates the strength of DM interaction. A comparative study between Cu(1,3-bdc) and Herbertsmithite within the same calculation framework is made in Sec. \ref{sec:compare}. Section \ref{sec:conclusion} looks back upon the previous experimental results and discusses remaining ambiguities.

\section{Calculation method} \label{sec:method}
The calculations are carried out using the VASP package \cite{VASP96}, which solves the DFT Hamiltonian self-consistently using the plane wave basis together with the projector augmented wave method \cite{VASPpaw99}. A plane-wave cutoff of 500 eV is enforced. The integration over the Brillouin zone is obtained on a $\Gamma$-centered $4\times4\times2$ k-mesh. The self-consistent iterations are converged to 0.1 meV precision of the total energy. We use the unit cell and lattice parameters determined by experimental X-ray diffraction \cite{JACS08CuBDC}. The Cu coordinations are automatically fixed by the hexagonal space group (P6$_3$/m) without forces. The light atoms of the 1,3-bdc ligands are fully relaxed until the forces are less than 0.01 eV/$\AA$.

Within this formalism, we first obtain the single-electron properties under the local density approximation (LDA) \cite{PZ81}, and down-fold the full band structure to a single-orbital hopping model. Then based on the \emph{ab initio} Wannier functions \cite{Vanderbil97MLWF}, we estimate the strength of various electron-electron interactions and determines the spin exchange. Further analysis is performed by using the generalized gradient approximation (GGA) \cite{PBE96} and +U functional \cite{Dudarev98PRB}.

We note that these two methods have successful applications in closely related transition-metal insulators \cite{Hybertsen90PRB,Anisimov97LDAU}. The Wannier function analysis has been used to explain the unexpected ferromagnetism in La$_4$Ba$_2$Cu$_2$O$_{10}$ \cite{WKu02WannierFM}. A recent DFT+U study has nicely reproduced the NN AFM exchange in Herbertsmithite \cite{DFT13Herbert}. Meanwhile, it is understood that first-principles predictions on sub-meV magnetic exchange are highly challenging, because either the exchange-correlation functional or the pseudopotential can easily introduce uncertainties at this scale. Therefore, as far as possible, we avoid drawing shaky conclusions that sensitively rely on the numerical precision. Instead, we proceed with evident and consistent numerical features of the material as a guide to construct reasonable physical understanding.

\section{Structural and single-electron properties} \label{sec:band}

\begin{figure}[ht]
\includegraphics[width=0.5\textwidth]{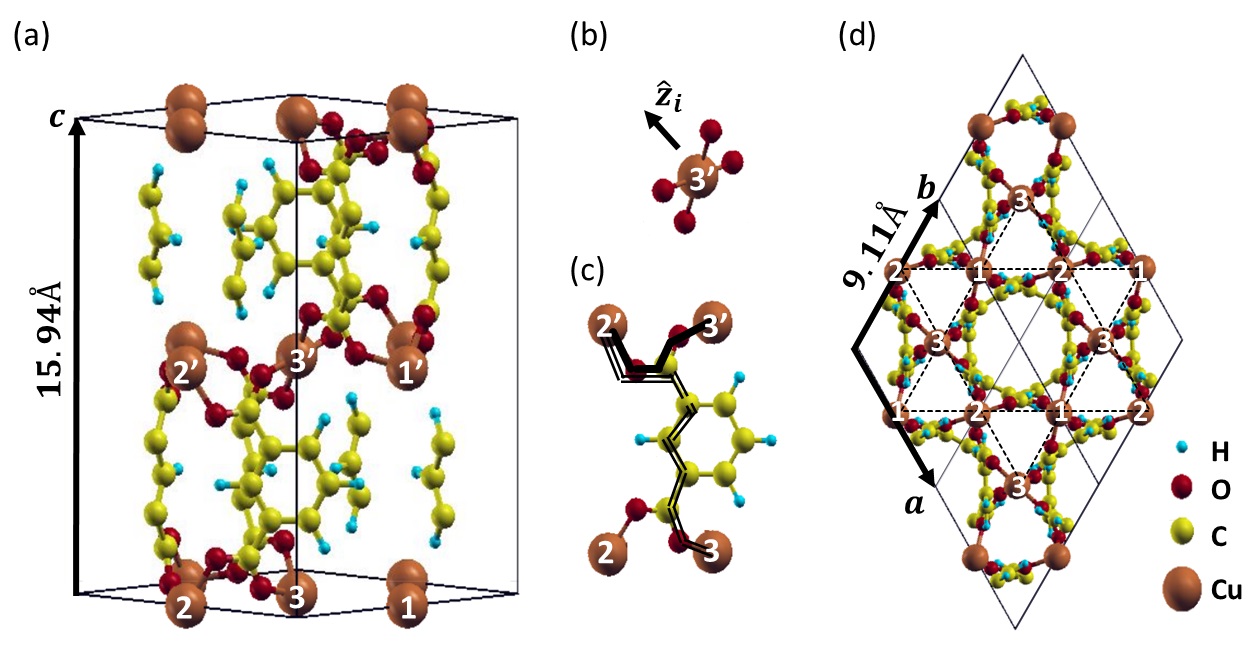}
\caption{\label{fig:structure}Atomic structure of Cu(1,3-bdc). The white numbers on Cu atoms label the three inequivalent sites of a kagom\'{e} plane. (a) Side view of the hexagonal unit cell. (b) Local planar coordination of Cu. (c) 1,3-bdc as a linker; the solid lines show the shortest intra-plane path between two Cu sites; the double lines show the shortest inter-plane path; (d) top view of the hexagonal unit cell; the dashed lines are guide to the eyes for the kagom\'{e} geometry formed by the Cu sites. }
\end{figure}

Figure 1 shows the atomic structure of Cu(1,3-bdc). The crystalized network forms a hexagonal lattice, containing two Cu kagom\'{e} planes per unit cell [Fig. 1(a)].  Each kagom\'{e} plane consists of three inequivalent Cu sites [Fig. 1(d)]. The local environment of the Cu atom is similar to that in the CuO$_2$ plane of cuprate superconductors: each Cu atom bonds with four O atoms in different 1,3-bdc ligands forming a local planar coordination [Fig. 1(b)]. Each 1,3-bdc ligand also bonds with four Cu ions via the two carboxyl groups, which mediates the intra-plane and inter-plane hopping [Fig. 1(c)]. The 1,3-bdc is in the $-2$ state, so it is clear that Cu has an oxidation number $+2$.

We start from the standard LDA (spinless) band calculation to understand the electronic properties. The result [Fig. 2(a)] shows six bands around the Fermi level, isolated from the other bands. This set of bands exhibits the typical feature of single-orbital hopping on a 2D kagom\'{e} lattice, i.e. a flat band and two dispersive bands with a linear crossing \cite{Zheng14CPB}. We will refer to these bands as the ``kagom\'{e} bands'' hereafter.  Recall that there are two Cu kagom\'{e} planes per unit cell, which give rise to two sets of kagom\'{e} bands. The flatness of the top bands suggests that except the NN hopping, all the other hopping processes are weak. The Fermi level crosses the middle of the six bands, which corresponds to half-filling of these states. It is well known that LDA can not properly describe the on-site Coulomb repulsion of $3d$ orbitals. Hence, the LDA calculation predicts a metallic phase.

\section{Wannier function analysis} \label{sec:wannier}

\begin{figure}[ht]
\includegraphics[width=0.51\textwidth]{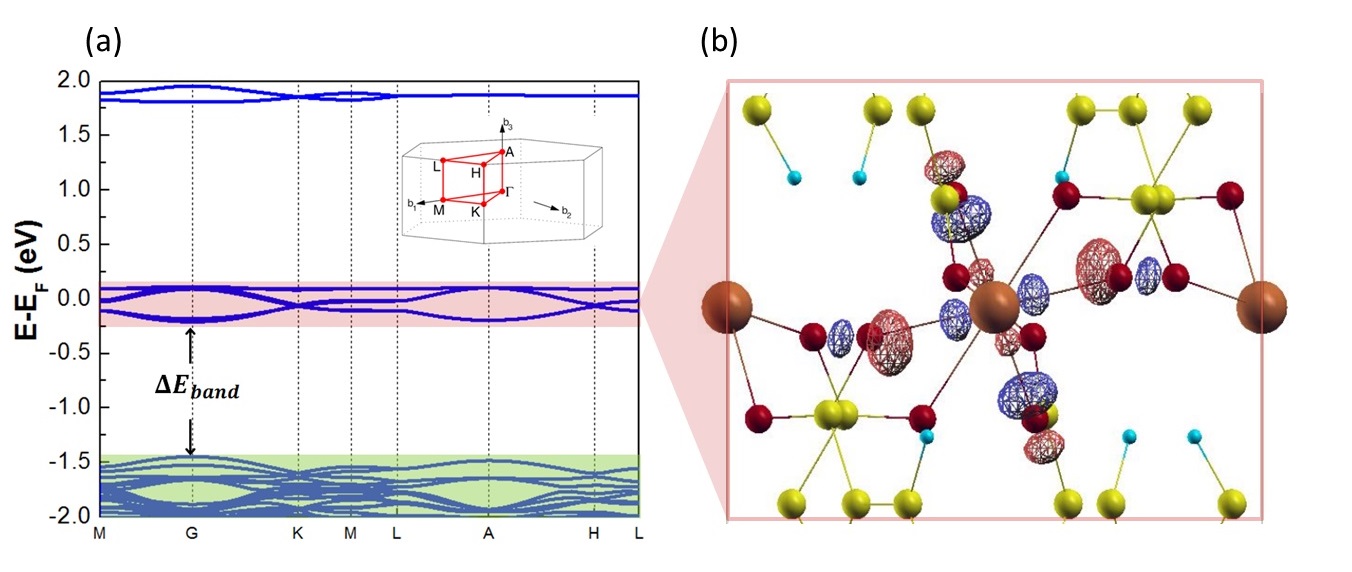}
\caption{\label{fig:singleelectron} (a) The single-electron (within LDA) band structure of Cu(1,3-bdc). The inset is the Brillouin zone and high-symmetry points of the hexagonal lattice. (b) Wannier function of the kagom\'{e} bands [red shaded in (a)] around the Fermi level. The wired surface plots the iso-value contour, and the color (red/blue) denotes the sign.  }
\end{figure}

The single-electron band structure can be understood by considering Cu$^{2+}$ ions under a planar crystal-field splitting, with a single $d_{x^2-y^2}$ at the top. The nine $d$-electrons in one Cu$^{2+}$ ion will fully occupy the bottom four orbitals, leaving an unpaired electron on $d_{x^2-y^2}$, which in the end reduces to a single-orbital degree of freedom around the Fermi level. The low-energy dynamics are primarily determined by this subspace, which is well defined in this case owing a large gap with other occupied bands [$\Delta E_{band}$ in Fig. \ref{fig:singleelectron}(a)]. Then, it is helpful to down-fold the full band structure into an effective single-orbital hopping model:
 \begin{eqnarray}\label{eq:hop}
H_{hop}=\sum_{i,j}t_{ij}c^\dag_i c_j,
\end{eqnarray}
where $i,j$ label the Cu site , and $t_{ij}$ is the hopping parameter between the two sites. To construct a quantitative basis, we perform Fourier transformation from the Bloch representation to the Wannier representation by using the Wannier90 code \cite{Vanderbilt08W90}. Figure 2(b) plots the spacial distribution of the maximally-localized Wannier function centered at one of the six Cu sites; the others are related via the crystal symmetry. The Wannier function takes the form of a hybridization between the Cu $d_{x^2-y^2}$ orbital and the O $p_\sigma$ orbital.

The hopping parameters between these Wannier functions can be rigorously calculated by performing the same Fourier transformation to the band structure. We list three leading terms in Tab. I: the NN hopping $t_1$, the 2nd largest in-plane hopping $t_2^{in}$ and the largest out-of-plane hopping $t_2^{out}$. The NN hopping $t_1$ is one order of magnitude larger than the other hopping terms, dominating the hopping dynamics. It is worth noting that $t_1$ has a nontrivial minus sign, which determines the position of the flat band. This sign cannot be simultaneously gauged away on the three Cu sites. When the electron circles the three sites, the minus sign leads to a $\pi$ Berry phase.

\begin{figure*}[ht]
\includegraphics[width=0.8\textwidth]{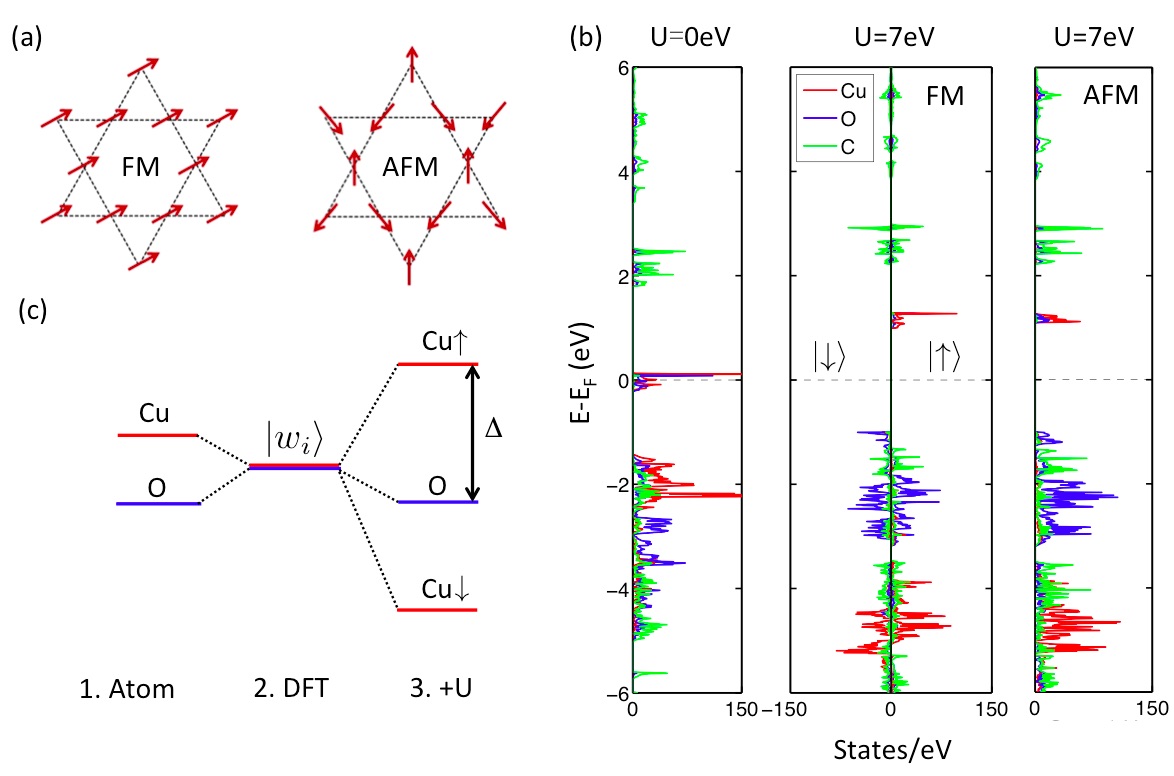}
\caption{\label{fig:DFT+U} (a) Two spin configurations for the DFT+U calculation (b) Projected density of states before and after the +U correction. (c) Schematic plot of the formation of energy states around the Fermi level}
\end{figure*}

\begin{table}[ht]
\caption{A comparison of key parameters for Cu(1,3-bdc) and Herbertsmithite derived from the DFT Wannier function analysis}
\begin{ruledtabular}
\begin{tabular}{cccc}
 & Cu(1,3-bdc) & Herbertsmithite & Ref. \\
  \hline
  & \multicolumn{2}{c}{Hopping (eV)} &\\
  $t_1$ & $-5.0\times10^{-2}$ & $1.8\times10^{-1}$ & Eq.(\ref{eq:hop})\\
  $t_2^{in}$ & $4.8\times10^{-3}$ & $2.3\times10^{-2}$ & Eq.(\ref{eq:hop})\\
  $t_2^{out}$ & $2.3\times10^{-3}$ & $3.7\times10^{-2}$ & Eq.(\ref{eq:hop})\\
  $\tilde{\lambda}$ & $1\times10^{-3}$ & $5\times10^{-3}$ & Eq.(\ref{eq:soc})\\
  \hline
  & \multicolumn{2}{c}{Bare interaction (eV)} & \\
  $U_0$ & $6.8$ & $6.4$ & Eq.(\ref{eq:Coulombintegral})  \\
  $U_1$ & $1.0$ & $4.1$ &  Eq.(\ref{eq:Coulombintegral}) \\
  $J_{ex}$ & $1.7\times10^{-4}$ & $5.6\times10^{-2}$ &  Eq.(\ref{eq:Coulombintegral})
\end{tabular}
\end{ruledtabular}
\end{table}

Even without information on the spin exchange, the single-electron properties shown above already suggest Cu(1,3-bdc) as an ideal $S=\frac{1}{2}$ kagom\'{e} model system: (a) the half-filled Wannier function gives rise to a half spin at each Cu site; (b) beyond the NN coupling, the additional perturbations, such as second neighbor and interplane couplings, are weak. In order to uncover the underlying spin exchange, we need to evaluate the many-body interactions between the Wannier functions not captured within LDA. The dominating interaction Hamiltonian contains three terms \cite{noteDMFT}:
\begin{eqnarray}\label{eq:int}
H_{int}=\tilde{U}_0\sum_i n_{i\downarrow}n_{i\uparrow}+\tilde{U}_1\sum_{\langle ij\rangle}n_i n_j\nonumber\\
+\tilde{J}\sum_{\langle ij\rangle\alpha}c_{i,\alpha}^\dag c_{j,-\alpha}^\dag c_{i,-\alpha}c_{j,\alpha},
\end{eqnarray}
where $\langle ij\rangle$ and $\alpha$ denotes the NN pairs and spin, respectively. $\tilde{U}_0$ is the on-site Hubbard repulsion, $\tilde{U}_1$ is the NN direct repulsion and $\tilde{J}_{ex}$ is the NN direct exchange. We explicitly include the inter-site direct exchange $\tilde{J}_{ex}$ to address the possibility of any ligand mediated Hund's coupling as phenomenologically formulated by the Goodenough-Kanamori rules \cite{Goodenough55,Kanamori59}. A similar Wannier function analysis has successfully explained the ferromagnetism in La$_4$Ba$_2$Cu$_2$O$_{10}$ \cite{WKu02WannierFM}.

We first evaluate the ``bare'' Coulomb integrals with respect to the Wannier functions as a 0th-order approximation to these interactions. The double-counting correction takes the form of an on-site chemical potential. For the half- filling case as what we are studying here, it amounts to a rigid energy shift. Since the screening effect is completely overlooked,  the bare values tend to overestimate the interaction strength. The on-site U responsible for the superexchange is further limited by the charge transfer gap as discussed later in Sec. V. The key point here is that this bare-parameter estimation sets the upper limit of the interaction-driven FM exchange and the lower limit of the kinetic-driven AFM superexchange. We are going to show that the FM exchange does not surpass the AFM superexchange even in such limit.

 The numerical results for the following integrals are listed in Tab. I: 
\begin{eqnarray}\label{eq:Coulombintegral}
U_0&=&\int d\textbf{r}d\textbf{r}'\frac{|w_i(\textbf{r})|^2|w_i(\textbf{r}')|^2}{|\textbf{r}-\textbf{r}'|}\nonumber\\
U_1&=&\int d\textbf{r}d\textbf{r}'\frac{|w_i(\textbf{r})|^2|w_j(\textbf{r}')|^2}{|\textbf{r}-\textbf{r}'|}\nonumber\\
J_{ex}&=&\int d\textbf{r}d\textbf{r}'\frac{w_i^*(\textbf{r})w_j(\textbf{r})w_j^*(\textbf{r}')w_i(\textbf{r}')}{|\textbf{r}-\textbf{r}'|},
\end{eqnarray}
in which $w_i$ is the Wannier function centered at site $i$.  The condition $U_0\gg t_1$ suggests that the electron model can be safely reduced to a Heisenberg spin model by the standard second-order perturbation:
\begin{eqnarray}
H_{spin}&=&J_1\sum_{\langle i,j\rangle} \textbf{S}_i\cdot \textbf{S}_j\label{eq:Heisenberg}\\
J_1&=&\frac{4t_1^2}{\tilde{U}_0-\tilde{U}_1}-2\tilde{J}_{ex},\label{eq:exchange}
\end{eqnarray}
in which $\textbf{S}_i$ is the spin $\frac{1}{2}$ operator at site $i$. Substituting the bare parameters into Eq.(\ref{eq:exchange}) gives $\frac{4t_1^2}{U_0-U_1}=1.72meV$ and $2J_{ex}=0.34meV$, and the net effective NN spin exchange is $J_1=1.38meV=16K$.  $J_1$ will be further pushed to the AFM side with screening.  Therefore, the 2K FM scenario is not supported. 

\section{DFT+U analysis}\label{sec:+U}
The DFT+U method incorporates at the Hartree-Fock level the strong correlation of localized atomic orbitals, and describes magnetism in an itinerant picture. In spite of its mean-field nature, this method has been proved to be an effective tool to provide sensible information on the electronic and magnetic properties of transition-metal insulators \cite{Anisimov97LDAU}.  The calculation involves two parameters $U$ and $J$, describing the average repulsion and Hund's exchange between the Cu $3d$ orbitals. Following the previous DFT+U calculations on cuprates and Herbertsmithite \cite{Anisimov97LDAU,DFT13Herbert}, we choose a variety of empirical $U$ ranging between 6 eV and 8 eV and $J=$1 eV. Note that these parameters should not be confused with those in Eq.(\ref{eq:Coulombintegral}), which refer to the down-folded Wannier functions. The DFT+U calculations represents an independent analysis based on a full description of the material, rather than a mean-field solution to Eq.(\ref{eq:hop})+Eq.(\ref{eq:int}).

The ground state of Cu(1,3-bdc) is expected to be a spin ordering state. We start from two spin configurations of the Cu kagom\'{e} plane as shown in Fig.3(a) to address the FM and AFM NN scenarios, respectively. After the self-consistent iteration is converged, the FM solution maintains the parallel spin configuration with only relaxed magnetic moment on Cu. The AFM solution slightly deviates from the perfect 120 degree configuration into an asymmetric 130, 130, 100 degree pattern. It is possible that spin configurations with lower energy exist in larger periodicity, but a comprehensive searching is computationally expensive. Notwithstanding, a comparison between these two typical configurations is sufficient to determine the type of the NN exchange. 

Figure 3(b) presents the projected density of states before and after the +U correction.  Without U, the Fermi level states are hybridized from the Cu and O states, which has been identified by the Wannier function analysis. Below the Fermi level, the nearest valence states also largely come from Cu. Hence, $\Delta E_{band}$ in Fig.2(a) reflects the size of crystal field splitting.  After the gap opening (with U turned on), the unoccupied band edge becomes Cu dominated; the occupied band edge becomes O dominated. The occupied Cu states are pushed deeper away from the Fermi level. For the FM configuration, the unoccupied band edge still exhibits the typical feature of kagome bands. For the AFM configuration, the density-of-states profile is renormalized due to the noncolinear spin structure.  According to Fig. 3(b), we draw a schematic plot of the electronic states around Fermi level in Fig. 3(c). Like in cuprates\cite{Hybertsen90PRB,ZhangRice88PRB}, the low energy excitation is between O and Cu, placing Cu(1,3-bdc) in the regime of charge-transfer insulator. Consequently, the AFM superexchange is mainly mediated by a transition state with double holes on O [See for example Eq.(4) in Ref. \cite{ZhangRice88PRB}]. The transition energy  $2\Delta\sim4eV$ plays the role of $\tilde{U}_0-\tilde{U}_1$ in Eq.(\ref{eq:exchange}). The corresponding AFM coupling strength is $4t_1^2/(2\Delta)\sim30K$, which agrees with the experimental value from Curie-Weiss fitting. A summary of the AFM $J_1$ values obtained from different methods are discussed in the Appendix. 

In Tab. II, we list the self-consistent total energy per unit cell ($E_T$), the energy gap around the Fermi level ($\Delta$), and the relaxed magnetic moment on Cu ($\mu_{Cu}$).  The results show several features robust to the variation of U-J parameters. Firstly, the AFM configuration is found to be lower in energy than the FM configuration. Secondly, DFT+U correctly reproduces the insulating phase, and the AFM configuration gives a gap slightly larger than the FM configuration. Thirdly, the calculated magnetic moment on Cu is similar to previous DFT+U results for cuprates \cite{Anisimov97LDAU}. The value is identical for different initial spin configurations, confirming the validity of an effective local spin model [Eq.(\ref{eq:Heisenberg})].  These evidences consistently show that $J_1$ is of the AFM type.

\begin{table}[ht]
\caption{Total energy per unit cell ($E_T$), energy gap around the Fermi level ($\Delta$), and relaxed magnetic moment on Cu ($\mu_{Cu}$) from the DFT+U calculation}
\begin{ruledtabular}
\begin{tabular}{cccc}
 U (eV) & 6 & 7 & 8 \\
 J (eV) & 1 & 1 & 1\\
  \hline
 $E_T^{FM}$ (eV) &  -696.710(9)  & -695.461(1) & -694.294(4)\\
  $E_T^{AFM} (eV)$ & -696.716(4) &  -695.465(6) &  -694.298(2) \\
  $\Delta E_T (K)$ & 63  &  52 & 44\\
    \hline
  $\Delta^{FM} $ (eV)& 1.8 & 2.0 & 2.2\\
  $\Delta^{AFM}$ (eV) & 1.9 & 2.1 & 2.3\\
  \hline
  $\mu_{Cu}^{FM}$ ($\mu_B$) & 0.66 & 0.68 & 0.71 \\
  $\mu_{Cu}^{AFM}$ ($\mu_B$) & 0.66 & 0.68&  0.71 \\
\end{tabular}
\end{ruledtabular}
\end{table}

\section{Effects of SOC} \label{sec:soc}
The SOC is responsible to various secondary spin anisotropic terms. These terms, especially the DM interaction \cite{Dzyl58,Moriya60} in quantum frustrated magnets, have attracted a lot of attention, due to their potentially important role in determining the ground state \cite{Singh07suscept,Lhuillier08schwinger,Lhuillier10PRBSchwinger}. By including SOC in the DFT(+U) Hamiltonian, we quantify its effects within the first-principles formalism.

\begin{figure}[ht]
\includegraphics[width=0.46\textwidth]{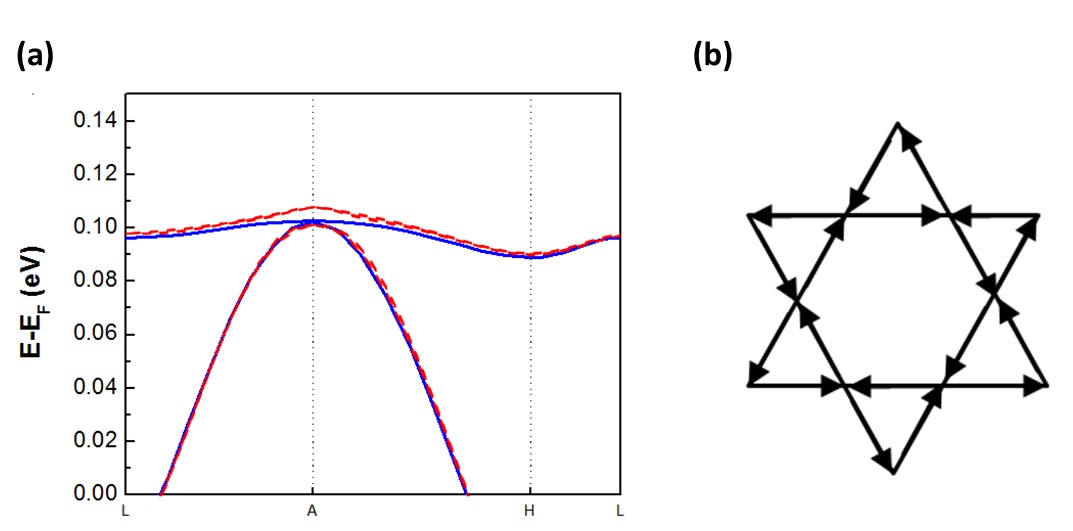}
\caption{\label{fig:soc} (a) The SOC-induced gap around $A$ point in the single-electron band structure. (Red dashed curve) with SOC; (blue solid curve) without SOC. (b) The hopping direction corresponding to a positive $\eta_{ij}$ in Eq.(\ref{eq:soc}).}
\end{figure}

For the kagome band around the Fermi level, the primary effect of SOC on the single-electron band structure is split the degeneracy at several k-points. In Fig.\ref{fig:soc}, we zoom in around the $A$ point to show a SOC-induced band gap of 6 meV. Note that this SOC-split gap is much smaller than the SOC constant of a free Cu atom. The reason is that the intra-atomic SOC manifests in the crystal-field-split $d_{x^2-y^2}$ subspace only through higher order perturbation, namely, the intra-atomic SOC first promotes the electron to underlying d-orbitals outside the subspace, and then the electron hops to the other site, altogether becoming an imaginary inter-atomic hopping \cite{Coffey91PRB}.  This gapping mechanism is recently an active topic, because of the associated nontrivial band topology \cite{PRB08BergmanTouch,PRL11Tang,Zheng14CPB}.  A simplified form of SOC on the kagom\'{e} lattice can be written down as \cite{PRL11Tang}:
\begin{eqnarray}\label{eq:soc}
H_{soc}=i\tilde{\lambda}\sum_{\langle i,j\rangle \alpha} \eta_{ij}c_{i\alpha}^\dag\sigma^z_{\alpha\alpha}c_{j\alpha},
\end{eqnarray}
in which $\tilde{\lambda}$ is the effective strength of SOC and $\sigma^z$ is the z-component Pauli matrix. $\eta_{ij}$ is a sign determined by the hopping direction: +, if following the arrows in Fig.\ref{fig:soc}(b); -, if opposite. This form of SOC conserves $S_z$; i.e., assuming the electric field on each site is in the 2D plane. Intuitively, it pins a nontrivial phase to the electrons when they hop around the lattice.  As shown later, this SOC leads to an out-of-plane DM interaction, which is typically the dominant spin anisotropic term.

By fitting the first-principles band splitting to Eq.(\ref{eq:soc}), the value of $\tilde{\lambda}$ can be determined to be 1 meV (Tab.I). We can now add $H_{soc}$ to $H_{hop}$, and do the second-order perturbation again with respect to $H_{int}$. Besides the isotropic Heisenberg exchange, the next largest interaction arises from the $\tilde{\lambda}t_1$ cross term:
\begin{eqnarray}
H_{DM}&=&\sum_{\langle i,j\rangle}D_{ij}^z(\textbf{S}_i\times\textbf{S}_j)_z\\
D_{ij}^z&=&\frac{8\tilde{\lambda}t_1}{\tilde{U}_0-\tilde{U}_1}\eta_{ij},
\end{eqnarray}
which is nothing but the out-of-plane DM interaction. Since $2J_{ex}\ll 4t_1^2/(\tilde{U}_0-\tilde{U}_1)$ in Cu(1,3-bdc), the ratio $|D_{ij}^z|/|J_1|$ is simply $2\tilde{\lambda}/|t_1|=1/25$. Taking $|J_1|\sim30K$, $|D_{ij}^z|$ is estimated to be of the order of 1K, comparable to the phase transition temperature $T_c$. The pseudo-dipole interaction $H_{aniso}=\sum_{\langle i,j\rangle}\Gamma_{\mu\nu}\textbf{S}_i^\mu\cdot\textbf{S}_j^\nu$ arises from the $\tilde{\lambda}^2$ terms, thus one more order smaller than the DM interaction.

Including SOC in the DFT+U calculation is found to have negligible effects. Both the self-consistent spin configuration and the energy difference are the same as described in Sec.\ref{sec:+U} without SOC. When we globally rotate the spins, the spin anisotropic energy can be observed showing an in-plane preference. The magnitude is less than 1 meV per unit cell. In summary, we conclude that the dominant role of SOC in Cu(1,3-bdc) is inducing a DM interaction between NN spins. Despite a weak magnitude, it can induce observable anisotropy as observed in the single-crystal measurement \cite{Keren14PRBCuBDC}, and is possibly related to the phase transition around 2K.

\section{Comparison with Herbertsmithite} \label{sec:compare}

Quoted as the end to the drought of QSL, Herbertsmithite has been extensively studied in the past few years \cite{Science08PALeeQSL}. Some of its properties are carefully determined experimentally, such as a dominant AFM NN coupling $J\sim$180K and a z-component DM interaction $D^z\sim1/10J$ \cite{JPSJ10ReviewHerbert}. Being the ``siblings'', it is informative to conduct a comparative study on these two materials within the same theoretical framework.

\begin{figure}[ht]
\includegraphics[width=0.5\textwidth]{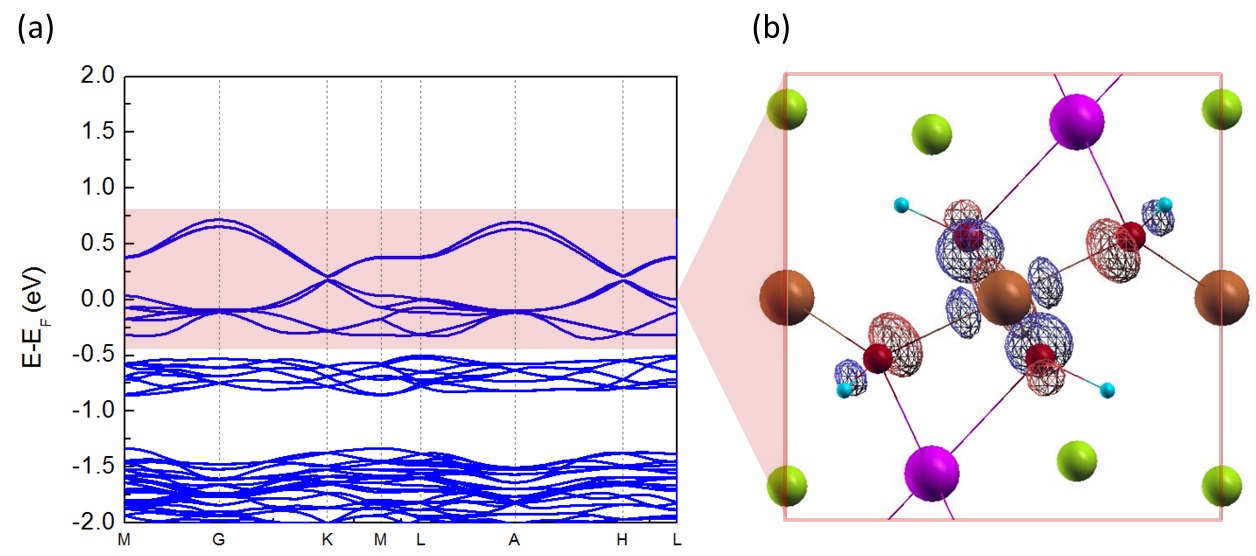}
\caption{\label{fig:herbert} (a) The single-electron band structure of Herbertsmithite.  (b) Wannier function of the kagom\'{e} bands [red shaded in (a)] around the Fermi level. The wired surface plots the iso-value contour, and the color (red/blue) denotes the sign. }
\end{figure}

With the rhombohedral (R-3m) space group, Herbertsmithite contains three Cu$^{+}$ kagom\'{e} planes per unit cell \cite{Nocera05Herbert}. The NN Cu atoms are bonded to one common O atom. The two kagom\'{e} planes are bridged by an O-Zn-O three-atom path. Hence, the kagom\'{e} planes are much more compact than in Cu(1,3-bdc).  We show the single-electron band structure and the Wannier function of Herbertsmithite in Fig.\ref{fig:herbert}. A quick comparison between Fig.\ref{fig:singleelectron} and Fig.\ref{fig:herbert} gives the following information:

(a) For Herbertsmithite, the bands around the Fermi level deviate from the ideal kagom\'{e} bands more significantly . Therefore, compared with Cu(1,3-bdc), the additional hopping terms beyond NN have larger magnitudes, as expected from the more compact structure.

(b) For Herbertsmithite, the band that can be traced back to the ideal flat mode resides on the bottom, opposite to the case in Cu(1,3-bdc). As discussed in Sec.\ref{sec:band}, the position of the flat band is determined by the nontrivial sign of $t_1$. The hopping sign comes from the overlap of the Wannier functions, which depends on the bonding pattern and distance.

(c) The Wannier function is similar as a consequence of the same local CuO$_4$ coordination. With this picture, Cu(1,3-bdc) can be roughly viewed as a loosely-packed Herbertsmithite.

To provide deeper insights, we list the parameters of Herbertsmithite calculated by the same Wannier function analysis along with Cu(1,3-bdc) for comparison (Tab.I). The hopping amplitudes are in general one order of magnitude larger in Herbertsmithite because of the shorter hopping path, yet the ratios, e.g. $|t_2|/|t_1|$ and $|\tilde{\lambda}|/|t_1|$ are roughly the same. With regards to the interactions, $U_0$ is almost the same, confirming the similarity of the Wannier functions; $U_1$ and $J_{ex}$ are larger in Herbertsmithite as expected.

\section{Conclusion and discussion} \label{sec:conclusion}
In summary, our DFT calculations suggest that Cu(1,3-bdc) closely reproduces the ideal $S=\frac{1}{2}$ kagom\'{e} AFM Heisenberg model. The relative strength of additional interaction terms with respect to the dominant NN AFM exchange is summarized as below.

(a) DM interaction: $\frac{\tilde{\lambda}}{t_1}\sim O(10^{-1})$.

(b) Next NN exchange and inter-plane exchange: $\frac{t_2^2}{t_1^2}\sim O(10^{-2})$.

(c) Pseudo-dipolar interaction: $\frac{\tilde{\lambda}^2}{t_1^2}\sim O(10^{-2})$.

Based on these numerical results, our overall understanding on previous experimental results is as follows. The local spin nature guarantees a nice Curie-Weiss behavior in the high-T range, so the Weiss constant $\theta=$-33K extracted from the high-T susceptibility [$\chi^{-1}(T)$] fitting should be respected, which defines a reliable $J_1$ energy. The deviation from the Curie-Weiss law occurs when $T$ is comparable or smaller than $J_1$ because of the breakdown of the molecular field picture, which makes the $\chi^{-1}(T)$ fitting no longer meaningful. Just as observed in Herbertsmithite, the downturn of $\chi^{-1}(T)$ (or equivalently upturn of $\chi(T)$) have complicated origins, leading to a false FM interpretation. The phase transition around $T_c=$2K may be associated with the additional secondary interactions, such as the DM interaction. The system undergoes an ordering transition, but due to the frustrated lattice and small spin value, quantum fluctuations persist as observed in $\mu$SR. The weak hysteresis after ordering is not from a fully-polarized FM order, but rather a canted Ne\'{e}l order.

It is known that the Schwinger boson mean-field theory (SBMFT) provides a satisfying description on the disorder-order transition of $S=\frac{1}{2}$ kagom\'{e} AFM Heisenberg model \cite{Sachdev92kagomedisorder}. Using the SBMFT language, the ordering transition is described as a Bose-Einstein condensation of spinons on the QSL ground state. Following this picture, Cu(1,3-bdc) can be viewed as a condensed QSL. The SBMFT predicts a flat spinon band at the top of the excitation spectrum \cite{PRB07ShenSBMT}. This property is in sharp contrast with the conventional spin-wave theory, which gives a flat magnon band at the bottom of the excitation spectrum \cite{PRB92Berlinsky}. Hence, the inelastic neutron scattering signal of Cu(1,3-bdc) can be very different from that of large-spin kagome AFM materials, such as iron jarosite, which have been found to agree with the spin-wave theory \cite{PRL06LeeFe}.

Unpublished neutron scattering data on Cu(1,3-bdc) has been orally reported, which suggests FM ordering below $T_c$  \cite{Robin14APS}. The primary evidence, however, appears to be a top flat mode observed in the inelastic spectrum, which is considered to be coincide with the spin-wave theory of a kagom\'{e} ferromagnet.  We note that this data may need to be re-examined carefully, because if the SBMFT describes Cu(1,3-bdc) correctly, the dynamic structure factor bears many features similar to the magnon branches of a kagom\'{e} ferromagnet, including a top flat peak (For reference, see Fig. 1a in Ref. \cite{Sachdev14NatPhys}). This so-called ``weather-vane'' mode \cite{Lhuillier10PRBSchwinger} has never been observed in materials before, and thus could be easily interpreted in a wrong way. A distinction between the FM and the AFM scenarios is the energy scale: if Cu(1,3-bdc) turned out to be a ferromagnet, the NN FM exchange is estimated to be $\sim$ 2K \cite{Keren14PRBCuBDC}, whereas the AFM scenario anticipates a NN exchange one order of magnitude larger. This energy scale will be unambiguously reflected by the width of the neutron scattering spectrum.

The remaining question is why Herbertsmithite can stay in a disorder phase, while Cu(1,3-bdc) is tuned into an ordering phase by seemingly weaker perturbations.  The only qualitative difference between these two materials shown by our calculations is the nontrivial NN hopping sign. However, this sign does not explicitly enter the Heisenberg model as well as the additional terms we have discussed, because they all arise from the secondary-order perturbation in terms of hopping. This sign will manifest in higher-order perturbations, and theoretically it is interesting to ask whether the sign of these higher-order terms select a specific ground state. Another obvious difference between Cu(1,3-bdc) and Herbertsmithite is that Cu(1,3-bdc) is intrinsically free from the Cu/Zn substitutional defects. Then the open possibility is that these defects indeed play an important role in the low-temperature magnetic properties.

\section{Acknowledgement}
We would like to thank X.-G. Wen, H. Yao , Y.-S. Wu and O. Starykh for helpful discussions. Z.L. is supported by Tsinghua University Initiative Scientific Research Program. F.L. and Z.L. acknowledge supports from DOE-BES (DE-FG02-03ER46027). Research at Perimeter Institute is supported by the Government of Canada through Industry Canada and by the Province of Ontario through the Ministry of Research.

\appendix
\section{On quantitative extraction of the spin exchange energy}
The traditional approach employed by the DFT community to extract the spin exchange energy is based on the total energy difference between different spin configuration. It is by no means a rigorous approach, especially for the highly-frustrated kagome AFM: on the one hand,  DFT(+U) is of the mean-field nature; on the other hand, the description on AFM within DFT(+U) is essentially based on a spin density wave picture that maps to a classical spin model. To proceed using this approach, we first consider the spin operators in Eq. (4) as ordinary vectors, i.e. mapping to the classical Heisenberg model.  Then,  it will be convenient for us to rewrite Eq. (4) into \cite{Book11FrustMag}:
\begin{eqnarray}
H_{spin}&=&J_1\sum_{\langle i,j\rangle} \textbf{S}_i\cdot \textbf{S}_j=\frac{J}{2}\sum_{\alpha}|\textbf{L}_\alpha|^2+const,
\end{eqnarray}
where $\textbf{L}_\alpha=\sum_{i\in\alpha}\textbf{S}_i$, and $\alpha$ us over the triangles formed by the NN $ij$ pairs of the sites. $\textbf{L}_\alpha$ is nothing but the total spin of each triangle. By assuming a classical-spin mapping, for the FM configuration $|L_\alpha|=\frac{3}{2}$; for the AFM configuration $|L_\alpha|=0$ [See Fig. 3(a) for reference]. Therefore,
\begin{eqnarray}
\Delta E_T=\frac{ZJ_1}{2}(\frac{3}{2})^2, 
\end{eqnarray}
where $Z=4$ is the number of triangles in each unit cell. Note that there are 2 kagome planes in the unit cell, and each plane contains 2 triangles. Consequently,  $J_1$ is $\frac{2}{9}\Delta E_T$, in the range of 10-14K based on $\Delta E_T$ in Tab. II.

Another approach to extract $J_1$ is based on Eq. (5) as we showed in the main text. It is commonly agreed that LDA band dispersion gives good estimation on $t_1$. The complexity lies in the rigorous extraction of the screened Coulomb parameters. The bare Coulomb integrals by using the Wannier function can be used as a rough estimation, but one should keep in mind that it in general underestimate the AFM superexchange due to the missing of screening. We obtain $J_1=16K$ using this method. 

For the specific case of Cu(1,3-bdc), we find that a more reasonable description of the AFM superexchange can be obtained by taking advantage of the the charge-transfer gap determined by the +U calculation. Being a charge-transfer insulator, the AFM superexchange is mainly mediated by a transition state with double holes on O [See for example Eq.(4) in Ref. \cite{ZhangRice88PRB} and Fig. 3(c)]. The transition energy  $2\Delta\sim4eV$ plays the role of $\tilde{U}_0-\tilde{U}_1$ in Eq.(\ref{eq:exchange}). The corresponding AFM coupling strength is $4t_1^2/(2\Delta)\sim30K$, which is in the best agreement with the experimental value from Curie-Weiss fitting.

\end{document}